\documentclass[twocolumn,showpacs,preprintnumbers,amsmath,amssymb]{revtex4}
\usepackage[english]{babel}

\usepackage{graphicx}

\newcommand{\be}[1]{ \begin{eqnarray} \mbox{$\label{#1}$} }

\newcommand{\ee}{\end{eqnarray}}
\newcommand{\pref}[1]{(\ref{#1})}

\newcounter{mycount}

\newcommand\ie {{\it i.e. }}
\newcommand\eg {{\it e.g. }}

\newcommand{\av}[1]{\langle #1\rangle}

\newcommand\etab {\bar\eta}
\newcommand\pop {{\mathcal P}}

\begin{document}

\title { Quantum Hall quasielectrons -  Abelian and non-Abelian}

\author{T.H. Hansson$^1$}
\author{M. Hermanns$^1$}
\author{N. Regnault$^3$}
\author{S. Viefers$^{1,2}$}
\affiliation{$^1$Department of Physics, Stockholm University, AlbaNova University Center, SE-106 91 Stockholm,
Sweden \\
$^2$Department of Physics, University of Oslo, Box 1048 Blindern, NO-0316 Oslo, Norway  \\
$^3$ Laboratoire Pierre Aigrain, Departement de Physique, ENS, CNRS, 24 rue Lhomond, 75005 Paris,  France}

\date{\today}

\begin{abstract} 

The quasiparticles in Quantum Hall liquids carry fractional charge and obey fractional quantum statistics. Of particular recent interest are those with non-Abelian statistics, since their braiding properties could in principle be used for robust coding of quantum information. 
There is already a good theoretical understanding of quasiholes both in Abelian and non-Abelian QH states. Here we develop conformal field theory methods that allow for an equally precise description of quasielectrons, and explicitly construct two- and four-quasielectron excitations of the non-Abelian Moore-Read state.

\end{abstract}

\pacs{73.43.Cd, 11.25.Hf, 71.10.Pm}

\maketitle

Quantum Hall (QH) quasiparticles are quite remarkable.  Not only do they carry fractional electric charge, they also obey fractional quantum statistics meaning that they are neither bosons nor fermions, but anyons. They are of two types -- equally important in experiments -- quasiholes and quasielectrons. While the theory for QH  quasi{\em holes} is very well developed, it has proven much harder to analyze quasi{\em electrons}. In this letter we develop methods based on conformal field theory  that allow us to describe quasielectrons in a way that parallels those used for the quasiholes. 

For most QH states the statistics is {\it Abelian} -- meaning that the many-anyon wave function only acquires  a phase factor $e^{i\theta } \neq \pm 1$ under quasiparticle exchange\cite{leinaas}. There are, however, more exotic possibilites; these can occur when the ground state in a sector with {\em fixed} quasiparticle positions is degenerate, and an exchange of particles $i$ and $j$ results in a unitary transformation $U_{ij}$ that mixes  the different components in the wave function. If the $U_{ij}$'s corresponding to different exchanges do not commute, the statistics is said to be {\em non-Abelian}\cite{mr,wen1}. 
Much of the present interest in this field is spurred by the recent theoretical proposals to use such non-Abelian anyons for topologically protected quantum computing\cite{comp}.  The formalism we develop here is general enough to be applied to both Abelian and non-Abelian QH states, and we present explicit results for quasielectron excitations of the most prominent candidate for a non-Abelian QH liquid, namely the Moore-Read pfaffian state that is believed to describe the QH state observed at filling fraction $\nu = 5/2$\cite{mr}.

Although the insight that QH quasiparticles are anyons originated from an analysis of the Laughlin wave function\cite{halphfs}, much of the subsequent understanding has been gained from studying the corresponding effective low-energy theories, where the statistical phase factors appear as Wilson loops formed by (one or several) non-dynamical  "statistical"  gauge fields endowed with a topological Chern-Simons action.
A deep connection between such Chern-Simons gauge theories, and the algebraic properties of the correlation functions - or conformal blocks - of certain conformal field theories (CFTs), was demonstrated in a paper by Witten in 1989\cite{witt}. Later Moore and Read applied these ideas to the QH effect, and conjectured that the electronic wave functions, both for the ground state and for quasihole states, could be expressed as conformal blocks of the relevant CFT.  They supported their conjecture by showing that many known wave functions are in this class, and they also proposed the by now famous Moore-Read state which supports quasiholes  with non-Abelian statistics. Other early work on the connection between the QH effect and CFT was done by Fubini\cite{fubini} and by Wen\cite{wen}. Subsequently, many of the important QH wave functions have been expressed in terms of conformal blocks\cite{rr,we1,we2,spinsing,exblocks}.

To be precise, the ground state wave function for $N$ electrons can be written either as a single correlator $ \av { V(z_1)\dots V(z_N)} $, where $V(z)$ is a local operator describing the electrons, or as an antisymmetrized sum of such correlators involving several different representations, $V_n(z)$, of the electrons. The Laughlin state\cite{laughlin}, the Moore-Read state, and the Read-Rezayi states\cite{rr} are examples of the former;  a state at level $n$ in the positive Jain series\cite{jainbook}, on the other hand, is an example of a state which requires $n$ different electron operators\cite{we1}. The quasihole states are obtained by inserting one or more local quasihole operators, $H(\eta)$, in the correlators. The resulting $N$-electron wave functions $\Psi(\eta_1\dots\eta_M; z_1\dots z_N) $ will depend parametrically on the quasihole positions $\eta_i$. The actual form of the operators $V_n$ and $H$ depends on the CFT in question, but it is severely constrained by general principles.
All operators have to be local with respect to the electron operators, meaning that the braiding of any two electrons and of any single  quasihole with any electron must be trivial.  Also, the fractional charge and statistics of the quasiholes are usually coded in the algebraic properties of the  operators. 
For instance, the fractional Abelian statistics of the quasihole operators pertinent to the $\nu=1/3$ Laughlin state can be read from the operator product expansion $H(\eta)H(\xi) \sim (\eta-\xi)^{1/3} $, while the non-Abelian statistics of the Moore-Read quasiholes is coded in the fusion rules of the Ising model.\footnote{
The Ising model consists of the operators \{1, $\psi$, $\sigma$\} with the fusion rules:
$\psi\times \psi = 1$,
$\psi\times \sigma=\sigma$ and
$\sigma\times \sigma= 1+\psi$.
}

Here we present a representation of quasielectron states that is very similar to that of quasiholes. We construct an operator $\pop (\eta)$, which, when inserted in correlators, gives holomorphic quasielectron wave functions. 
This operator has the expected charge and conformal dimension, and  while not local, it is  quasi-local in the sense of having support on a region of the size of a magnetic length around the position $\eta$.  
We will give an explicit expression for the $\pop$ operator and outline how it can be used to obtain explicit wave functions for localized  two- and four-quasielectron excitations of the Moore-Read state, and we present numerical evidence for the charge of the quasielectron to be $-e/4$. To our knowledge no such wave functions have been published before.

Since quasiholes and quasielectrons can annihilate each other, one might expect that inserting the inverse of a quasihole operator, $H^{-1}(\eta)$ in a correlator, would yield  good quasielectron wave functions. Indeed, the charge and braiding properties of this operator are precisely what is expected for a quasielectron. Nevertheless, this simple approach does not work, since the resulting wave functions are not holomorphic 
but typically have singularities $\sim 1/(z_i - \eta)$, where $z_i$ is an electron coordinate.
As explained in \cite{we1}, the reason for this is the Pauli principle --- the operator $H^{-1}(\eta)$  "pulls" the electron liquid toward the position $\eta$, but there is a finite probability that an electron is  already there. 
At a heuristic level this difficulty can be overcome by an {\em ad hoc} projection onto the lowest Landau level, but a more precise and appealing approach  was recently proposed in Refs. \onlinecite{we1}, where the basic  idea was to  {\em shrink the correlation holes} around the electrons in the vicinity of the position $\eta$. 
%
Based on this insight, the  operator for a charge $-q_h e$ quasielectron is given by 
\be{qpspec}
\mathcal P(\eta) &=& \int d^2w \, e^{-\frac{1}{4\ell^2}q_h( |w|^2+|\eta|^2 - 2\bar\eta w ) } \left( H^{-1} \, \bar \partial J\right)_n(w)  \, ,
\ee
where $\left(\dots\right)_n$  stands for  a generalized normal ordering needed to define the products of operators, and 
$ J(z)$ is the holomorphic $U(1)$ current related to the electric charge. 
Since this current is conserved, its divergence has support only at the positions of the charges. Thus, when computing correlators of the type 
$\av { \mathcal P (\eta) \, \prod_i V(z_i)}  $, one obtains  delta functions, $\delta^2(w-z_i)$, at the positions $z_i$ of the electrons. Since $|w|^2 +|\eta|^2- 2\bar\eta w =  |w-\eta |^2 - (w\bar \eta-\bar w \eta)$ where the last term is purely imaginary, 
the exponential factor equals $e^{-\frac{1}{4\ell^2}q_h |w - \eta|^2  } $ up to a pure phase.  
Thus, the operator is exponentially localized around the position $\eta$.  The $w$-integral will give terms $\sim \left(H^{-1} V \right)_n(z_i)$, which amounts to putting an inverse quasihole of electric charge $-q_h e$ on top of an electron at point $z_i$. This is the mathematical meaning of  "shrinking the correlation hole around the electrons".  
The normal ordering,  which in the simple case of the Laughlin states coincides with the one conventionally used in CFT,\footnote{
See, \eg   \textsection  6.5 in ref. \onlinecite{gula}.} is given by
\be{deffus}
\left( A B \right)_n  (w) \equiv   \oint_w dy\, T(y)\oint_w dz \,A(z) B(w) \, ,
\ee
where $T(y)$ is the holomorphic energy momentum tensor of the CFT in question, and both the $y$ and $z$ contours are infinitesimal circles around the point $w$, the former enclosing the latter. This ordering prescription guarantees that $\pop$ has the same charge and conformal dimension as $H^{-1}$. Also note that the $y$-integration in \pref{deffus} amounts to taking the first descendant of the operator to the right.


Here we must explain an important point that might at first look purely technical, but is crucial for our construction, and also highlights an important conceptual issue in the description of anyonic quasiparticles.  The problem we face is the following:
Since the product $\left(H^{-1} V \right)_n(z_i)$ in \pref{qpspec} amounts to fusing the electron, which is a fermion, with a quasihole, which is an anyon, the corresponding multi-quasielectron  wave function will have unacceptable cuts in the {\em electron} coordinates.

To explain how to overcome this difficulty, we again consider the simpler case of  two quasielectrons in the 1/3 Laughlin state. Here the electron and quasihole operators are given by $V(z) = e^{i \sqrt 3 \varphi(z)}$ and $H(z) = e^{i \varphi(z)/ {\sqrt 3}}$, respectively, 
where $\varphi$ is an appropriately normalized scalar field. 
%
$V$ is fermionic since $V(z)V(w)\sim (z-w)^3$, and while $H$ is anyonic, it still has trivial braiding with the $V$'s as seen from the expansion $V(z)H(\eta)\sim (z-\eta)$. 
Thus the  correlator $\av{ H(\eta_1)H(\eta_2)V(z_1)\dots V(z_N)} $ gives a wave function that is holomorphic in the electron coordinates $z_i$, but with a normalization factor of the form $\eta_{12}^{1/3}=(\eta_1-\eta_2)^{1/3}$ characteristic of $\theta = \pi/3$ statistics of the quasiholes. This is, however, not the full story, since in a physical exchange process there can also be a Berry phase, and only the sum of the "monodromy" phase in $\eta_{12}^{1/3}$ and the Berry phase has a physical meaning.  
For the two-quasihole Laughlin state, constructed as above, this Berry phase can be shown to be zero. If we chose to remove the phase  $(\eta_{12}/\bar\eta_{12})^{1/6}$ in the wave function, the monodromy is zero, and the statistical phase must instead be extracted from the Berry phase. 
The point is, that as long as we  do not change their correlation properties vis-$\grave {\rm a}$-vis the electrons,  we are free to change the monodromy of the quasiholes at will, since this leaves the wave functions unchanged up to an unimportant phase factor\cite{halphi}. 
In the CFT approach, we can achieve this by introducing auxiliary scalar fields $\chi_i$. For instance, 
$H_b(\eta)      = e^{i \varphi(\eta)/ {\sqrt 3} -i  \sqrt{5/3 }\chi(\eta)   }  $ is a boson version of the charge 1/3 Laughlin hole, but 
$H_b$ and $H$ clearly have the same correlations with the electrons, since  $V$ is independent of $\chi$.

While for quasiholes it is entirely a matter of taste which representation to use, it is of utter importance for the quasielectrons. Using $H$ in \pref{qpspec} yields wave functions  containing factors $(z_i - z_j)^{4/3}$ reflecting the anyonic nature of $H$, while using $H_b$, instead gives the holomorphic factors $(z_i - z_j)^3$. 
In general, to avoid branch cuts, we need to insert a quasihole in \eqref{qpspec} with trivial monodromy, \ie having either fermionic or bosonic statistics, and the resulting electronic wave functions do differ.
Our construction is thus not unique in that the charge profiles can vary, but still the total charge, and the fractional statistics, remain unaffected by such short distance effects\cite{next}. The charge profile of an actual pinned quasielectron will depend on the details both of the interaction and the pinning potential.

Forming $ \mathcal P_b$ by inserting  the bosonic quasihole $H_b$ in \pref{qpspec},  and evaluating the correlator
$\av{\mathcal P_b (\eta_1) \mathcal P_b (\eta_2) V(z_1)\dots V(z_N) } $, exactly reproduces the Jain composite fermion 
wave function for  two localized quasielectrons\cite{jainbook}. This result directly generalizes to many-quasielectron excitations in all  the hierarchy states obtained by quasielectron condensation that were discussed in Ref. \onlinecite{we2}. 
In order to keep the discussion short, we have glossed over a number of  details that are of no conceptual importance, such as \eg the need to introduce a neutralizing background charge\cite{next}.

We now turn to  the Moore-Read (MR) state, and discuss its  quasielectron states in some detail.
 Usually the MR electron operator is expressed as a product of a bosonic exponential $e^{i\sqrt 2\varphi (z)}$, where $\varphi$ is a scalar field, and a Majorana fermion $\psi (z)$, and the non-Abelian statistics can be read from the monodromies of the pertinent conformal blocks.\footnote{
This conclusion relies on the assumption that in this formulation there is no Berry phase contribution to the statistics\cite{4qh,berryphas}.
}
For our purpose it is important to instead use an alternative form using only scalar fields, where electrons and quasiholes are described by
\be{mrop}
V(z) &=& \cos(\phi(z)) e^{i\sqrt 2 \varphi(z)}  \nonumber   \\
H_{\pm}(w)  &= & e^{\pm \frac{i}{2} \phi(w)+i\frac{1}{\sqrt{8}}\varphi(w)}  e^{-i\sqrt{\frac{3}{8}}\chi_1(w)\mp \frac{i}{2}\chi_2(w)} \, , \label{hole}
\ee
respectively. 
The electric charge is related only to $\varphi$, \ie $J=\frac{i}{\sqrt{2}}\partial \varphi$ in \eqref{qpspec}, while $\phi$ is purely topological and codes the non-Abelian statistics. The fields  $\chi_1$ and $\chi_2$ are introduced to make the statistics of the quasiholes trivial, in analogy to the Abelian case.

In the Majorana representation of the MR state, there is a unique quasihole operator $\sigma (\eta)e^{i \varphi (\eta)/\sqrt 8 }$, and the groundstate degeneracy in sectors with four or more quasiholes at fixed positions, arises because of the non trivial fusion rules of the Ising spin operators $\sigma (\eta)$. In the boson description \pref{hole}, 
there are three\footnote{ 
The $Z_2$ symmetry related to an overall change between  + and - reduces the naive counting with a factor 2. }
nonequivalent orderings $\av{ H_+(\eta_1)     H_+(\eta_2)   H_-(\eta_3)   H_-(\eta_4)   }$, $\av{ H_+   H_-H_+ H_-}$,  and  $\av{ H_+  H_-H_-H_+ }$ of four quasihole operators.  
One can show that only two of these give linearly independent correlation functions, and these span exactly the space of degenerate four-quasihole states found in Ref. \onlinecite{4qh}.
Substituting \pref{hole} in \pref{qpspec} (with $q_h=1/4$), and using the $T$   appropriate for a multicomponent boson{\footnote {T(z) is the sum of all four holomorphic energy-momentum tensors. }} in \pref{deffus}, we get a unique candidate for the  two-quasielectron wave function
\be{two}
\Psi =\langle \mathcal{P}_+(\eta_1)\mathcal{P}_-(\eta_2)\prod_{i=1}^NV(z_i) \rangle. \nonumber
\ee
In the figure we show the charge distribution for a state of  two   quasielectrons at antipodal positions on a sphere. It nicely illustrates that our wave function exhibits the expected separation of two localized $e/4$ charges. 
\begin{figure}[h!]
\begin{center}
\resizebox{!}{68mm}{\includegraphics{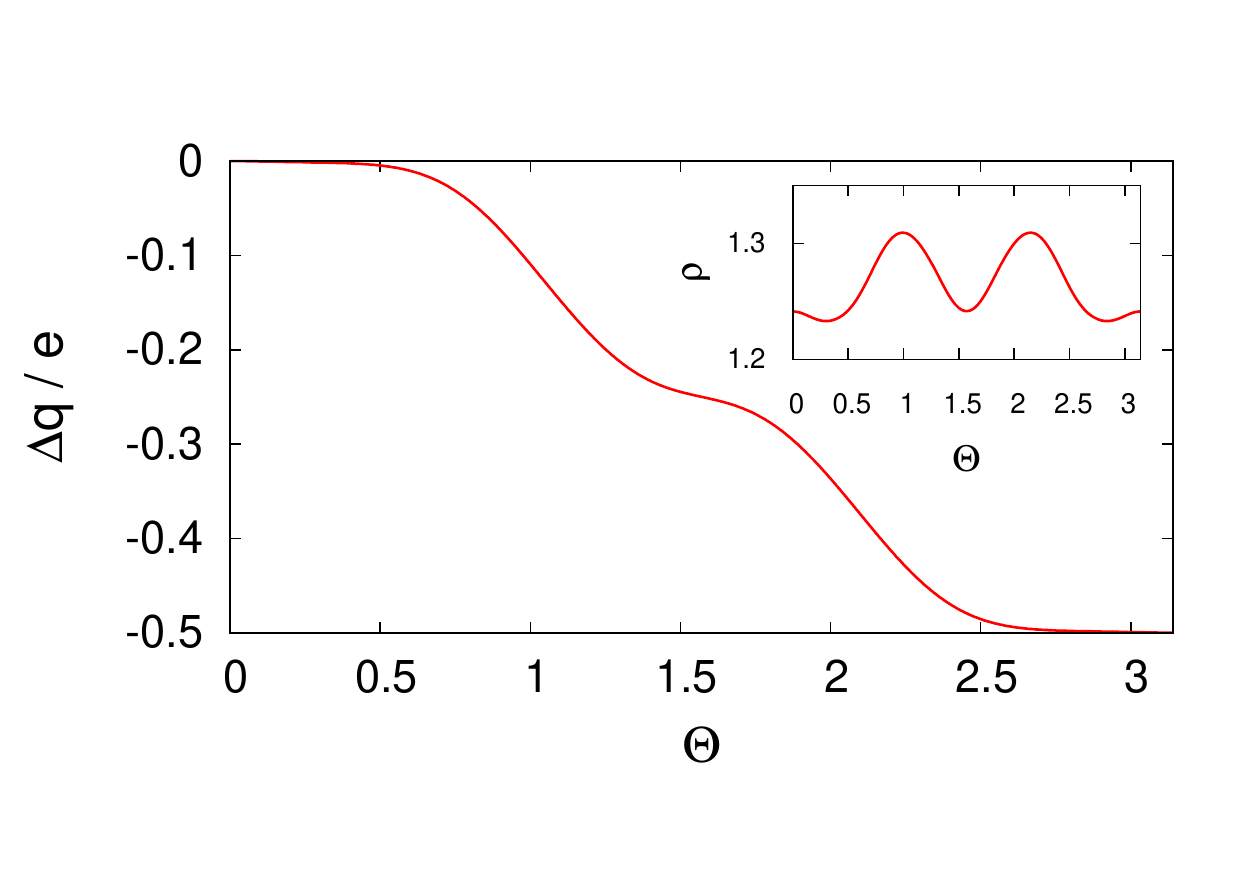}}
\end{center}\caption{\textit{{\small 
{\it Main plot} : excess charge $\Delta q$ of the two MR quasielectrons for $N=16$ particles on the sphere geometry ($\theta$ being the polar angle). The two quasielectrons with charge $e/4$ at $\nu=1/2$ are placed at the antipodes of a sphere corresponding to $\theta =0$ and $\theta = \pi$. The calculation has been done using Monte-Carlo integration. {\it Inset} : corresponding density profil.
}}}\label{frac}
\end{figure}
In  the case of four quasielectrons, just as for four quasiholes, the three nonequivalent arrangements of two $H_+$ and two $H_-$ lead to three distinct functions, $\Psi_{(1,2)}$, $\Psi_{(1,3)} $ and $\Psi_{(1,4)} $, two of which we  have shown to be linearly independent, where the indices denote the positions of the $\cal{P}_+$'s, for example
\be{four}
\Psi_{(1,3)} = 
\langle \mathcal{P}_+(\eta_1)\mathcal{P}_-(\eta_2) \mathcal{P}_+(\eta_3)\mathcal{P}_-(\eta_4) \prod_{i=1}^NV(z_i)  \rangle \, .\nonumber
\ee
These are our candidates for the two conformal blocks describing the four-quasielectron excitations of the Moore-Read state. The explicit expressions, which are lengthy and not very illuminating, will be given elsewhere\cite{next}.

Since our formalism allows us to write well-defined states of any number of localized, non-Abelian quasiparticles, it provides a starting point for constructing condensates of such particles using the idea of a hierarchy.  Recall that according to the  original proposals\cite{halphi,haldhi}, the hierarchical wavefunctions at level $n+1$ are coherent superpositions of quasiparticle excitations in the level $n$  parent state.
\be{hiwf}
\Psi_{n+1} ( \vec r_1\dots \vec r_{N}) &= &  \int  d^2\vec R_1 \dots   \int  d^2\vec R_M    \, \Phi^\star(\vec R_1 \dots \vec R_M )  \nonumber \\
&& \Psi_n (\vec R_1 \dots \vec R_M ; \vec r_1 \dots \vec r_{N}) 
\ee
where $\Phi(\vec R_1 \dots \vec R_M ) $ is a suitably chosen  "pseudo wave function" for the quasiparticles. Let us first illustrate with the simplest case of an Abelian quasielectron condensate in the $\nu=1/3$ Laughlin state.

Using the boson form $\mathcal P_b$ of the operator \pref{qpspec}, we get the following multi-quasielectron state,
\be{multqasi}
 \Psi (\vec R_1 \dots \vec R_{N/2} ; \vec r_1 \dots \vec r_{N})  &=&  \\
  \av {    \mathcal P_b(\eta_1)  \dots   \mathcal P_b(\eta_{N/2})    V(z_1)  &\dots&  V(z_{N})   }  \nonumber
 \ee
where $\eta = X + iY$.  Since $\mathcal P_b$ is bosonic, we must use a fully symmetric pseudo wave function $\Phi$. 
The natural choice describing $N/2$ charge $q = -e/m$ bosons in a magnetic field is 
\be{pseubos}
\Phi_k(\vec R_1 \dots \vec R_{N/2} ) = \prod_{i<j}^{N/2} (\etab_i - \etab_j)^{2k}  e^{-\frac 1 {4m\ell^2} \sum_{i=1}^{N/2} |\eta_i|^2}  \, . 
\ee 
The only $\etab$-dependence in $\mathcal P(\eta)$ is in the exponential factor $e^{\frac{1}{2m\ell^2} \bar\eta w } $ that will turn into $e^{\frac{1}{2m\ell^2} \bar\eta z_i } $  after the $w$-integration. Combined with the gaussian factors from \eqref{qpspec} and \pref{pseubos}, this yields 
$
\delta (\eta - z_i) \sim e^{-\frac 1 {2m\ell^2}( |\eta|^2 - \bar\eta z_i )}
$
which is nothing but a holomorphic delta function. Since the non-exponential part of $\Phi_k$ is holomorphic in $\etab$, the integrand in \pref{hiwf} is holomorphic, and thus the integrals can be done trivially because of the delta functions. 

It is now straightforward to show that the resulting wave function is {\em exactly} the $\nu = 2/5$ Jain composite fermion wave function written in terms of CFT correlators\cite{we1}. Similar considerations will, {\it mutatis mutandis},  apply to all states in the positive Jain series, and we believe that this  affirms that the CF wave functions can be seen as  hierarchical,
as already  pointed out by Read\cite{read90} and others\cite{TThier,we1,we2}.

It should now be clear that eq. \pref{multqasi} can also be applied to the Moore-Read state, and using the quasiholes \pref{hole} in $\pop$ we can use an appropriate ansatz for the pseudo wave function to construct candidate wave functions for condensates of non-Abelian quasielectrons, that  might give rise to an interesting family of hierarchical non-Abelian states.\footnote{
There is an earlier proposal for non-Abelian hierarchical states  based on condensing {\em Abelian} quasiparticles \cite{exblocks}.}
A similar approach to the parafermionic states also seems possible.

 In conclusion, we have given the general expression \pref{qpspec} for a quasilocal operator $\cal P$ that creates a localized quasielectron when inserted in CFT correlators. Using this we have explicitly constructed two- and four-quasielectron wave functions for the non-Abelian MR state, and we presented numerical support for the claim that the former indeed describes two localized charge $-e/4$ particles.
Finally, we offered a conjecture about the possibility of forming condensates of non-Abelian quasiparticles, which might give rise to a novel class of non-Abelian QH states.

After the completion of this manuscript we received a preprint by Bernevig and Haldane, that also discusses non-Abelian quasielectrons\cite{domkraft}.

We thank  Eddy Ardonne and Anders Karlhede  for many stimulating discussions and insightful remarks.
This work was supported by the Swedish Research Council, the Norwegian Reserach Council, and by NordForsk.

\end{document}